\documentclass{aastex}
\pdfoutput=1
\usepackage{graphicx}
\usepackage{natbib}
\newcommand{\et}{{\it et al.}}
\newcommand{\kms}{${\rm km\,s}^{-1}$}

\newcommand{\simless}{\mathbin{\lower 3pt\hbox
     {$\rlap{\raise 5pt\hbox{$\char'074$}}\mathchar"7218$}}} 
\newcommand{\simgreat}{\mathbin{\lower 3pt\hbox
     {$\rlap{\raise 5pt\hbox{$\char'076$}}\mathchar"7218$}}} 
\slugcomment{\it Submitted to the Astrophys. J. Letters; revised version of  08/13/13}
\shorttitle{Rule \et}
\shortauthors{Recombination lines from galaxies}

\begin{document}
\bibliographystyle{apj}
\baselineskip 24pt

\title{HIGH-$n$ HYDROGEN RECOMBINATION LINES FROM THE FIRST GALAXIES}
\author{E. Rule}
\affil {Johns Hopkins University \& Maria Mitchell Observatory, 4 Vestal Street, Nantucket, MA 02554}
\author{A. Loeb}
\affil {Harvard-Smithsonian Center for Astrophysics, 60 Garden Street, Cambridge, MA 02138}
\author {V.S. Strelnitski}
\affil {Maria Mitchell Observatory, 4 Vestal Street, Nantucket, MA 02554}

\begin{abstract}

We investigate the prospects of blind and targeted searches in the radio domain (10~MHz to 1~THz) for high-$n$ hydrogen recombination lines from the first generation of galaxies, at $z\simless 10$. The expected optically thin spontaneous $\alpha$-line luminosities are calculated as a function of the absolute AB magnitude of a galaxy at 1500\AA. For a blind search, semi-empirical luminosity functions are used to calculate the number of galaxies whose expected flux densities exceed an assumed detectability threshold. Plots of the minimum sky area, within which at least one detectable galaxy is expected at a given observing frequency, in the fiducial instantaneous passband of $10^4\,$\kms, allow to assess the blind search time necessary for detection by a given facility. We show that the chances for detection are the highest in the mm and submm domains, but finding spontaneous emission in a blind search, especially from redshifts $z\gg 1$, is a challenge even with powerful facilities, such as ALMA and SKA. The probability of success is higher for a targeted search of lines with principal quantum number $n\sim 10$ in Lyman-break galaxies amplified by gravitational lensing. Detection of more than one hydrogen line in such a galaxy will allow for line identification and a precise determination of the galaxy's redshift.
\end{abstract}

\section{INTRODUCTION}

Detection of spectral lines from the first generation of galaxies responsible for the re-ionization of the Universe is one of the primary goals of observational cosmology \citep{loeb12}. Most of the galaxies currently known at redhifts $z$=8-10 were detected photometrically \citep{2011ApJ...737...90B,2012ApJ...758...93F}, and have significant uncertainties in their redshift identification. More accurate determination of the redshift by spectroscopic identification would allow for probing the evolution of cosmological  structure on small scales at early cosmic times and it would shed light on the sources for the reionization of hydrogen. Detection of Lyman-$\alpha$ and other low-$n$ Rydberg lines of hydrogen is challenging and was proven practical so far mainly at $z < 7$ \citep{2010ApJ...723..869O}. Complementary information can be obtained from 21-cm surveys \citep{2012RPPh...75h6901P}.

Here we investigate the possibility of detecting high-$n$ hydrogen recombination lines from the large H$^+$ regions expected around the hot, massive stars in the first generation of galaxies. Extragalactic hydrogen recombination radio lines in cm and mm domains have been detected in more than a dozen of starburst  and Seifert galaxies [see chapter 3.5 in \citet{2002ASSL..282.....G} for a review]. The farthest galaxy with positive detections of radio recombination lines is Arp 220, at $z = 0.018$, with flux densities up to $\sim 100\,$ mJy in mm lines \citep{2000ApJ...537..613A}. The typical width of the emission features in extragalactic radio recombination line spectra is $\sim 100\;$\kms. These results were explained by spontaneous or amplified (maser) radiation from a multitude of H$^+$ regions of various sizes, from $\sim 0.1\,$pc to a few pc.

Although the sensitivity of radio telescopes and interferometers is growing rapidly, and ${\rm\mu}$Jy and even nJy thresholds of detectability are being discussed for the new and upcoming facilities, the jump from $z \sim 0.02$ to $z\sim 10$, corresponding to an increase of luminosity distance by three orders of magnitude (from $\sim 10^2$ to $\sim 10^5\,$Mpc) is challenging. In \S~2 we describe our method of calculation. We derive a simple analytical expression relating the expected flux densities in H$n\alpha$ lines, spontaneously emitted from an optically thin medium, with the galaxy's absolute AB magnitude at 1500\AA. Using the parameters of the monochromatic 1500\AA\ luminosity functions derived by \citet{2013ApJ...768L..37T} for a discrete set of redshifts in the range $z \le 10$, we develop empirical equations providing the Schechter parameters for any $z$ in this range. In \S~3, we present and discuss the results of the calculations. Using the numerical code described in \S~2, we find the number of galaxies whose flux density in at least one H$n\alpha$ line surpasses a given threshold of detectability at a given observing frequency, within a fiducial instantaneous passband. This allows us to consider the prospects of a blind search for these lines (\S~3.1). In \S~3.2, we show that, at present, the highest probability of detecting hydrogen recombination lines from galaxies with $z\simgreat 6-7$ is for Lyman-break galaxies amplified by gravitaional lensing. In \S~3.3 we briefly consider several effects that may influence our estimates of the flux densities and show that probably none of them can change our predictions by more than an order of magnitude. Our conclusions are summarized in \S~4.

\section{METHOD OF CALCULATION}

Our goal is to estimate the expected flux densities in high-$n$ hydrogen recombination lines from remote galaxies in the radio domain (from 10 MHz to 1 THz), in order to see whether a blind or targeted search for such lines is practical with existing or forthcoming facilities. Specifically, for a blind search, we want to determine the minimum sky area, $\Omega_{\Delta\nu}(\nu_0$), within which one can expect to find at least one galaxy satisfying the following three conditions: (1) its redshift $z\leq 10$; (2) its spontaneous, optically thin radiation in at least one hydrogen recombination line falls within a given instantaneous passband $\Delta\nu_0$ around the chosen observing frequency $\nu_0$; and (3) the expected flux density of this radiation exceeds a certain threshold, $f_{th}$. We limit ourselves to the strongest, H$n\alpha$, lines, i.e. the lines due to $(n+1)\rightarrow n$ transitions, where $n$ is the principal quantum number.

The redshift at which the line has to be emitted in order for it to be red-shifted to the observed frequency $\nu_0$ is:
\begin{equation}
z_e\, (\nu_0,n)= \frac{\nu - \nu_0}{\nu_0} = \frac{c\,{\rm R_H}}{\nu_0}\,\biggl [\frac{1}{n^2}-\frac{1}{(n+1)^2}\biggr ]-1\,,
\end{equation}
where $\nu$ is the rest frequency of the transition and ${\rm R_H = 1.0968\times 10^5\, cm^{-1}}$ is the Rydberg number for hydrogen. Differentiating equation~(1) gives:
\begin{equation}
{\rm d}z_e = -\nu\, \frac{{\rm d}\nu_0}{\nu_0^2}.
\end{equation}
Integration over the passband,
\begin{equation}
\Delta \nu_0 = \nu_0\,\frac{\Delta v_0}{c},
\end{equation}
gives the redshift interval corresponding to the width of the passband:
\begin{equation}
\vert \Delta z_e \vert = \nu\, \bigg |\frac{1}{\nu_0 + 0.5\Delta\nu_0} - \frac{1}{\nu_0 - 0.5\Delta\nu_0}\bigg | = \nu\,\biggl ( \frac {\Delta\nu_0}{\nu_0^2 - 0.25\,\Delta\nu_0^2}\biggr ) \approx \frac{\nu}{\nu_0}\,\frac{\Delta \nu_0}{\nu_0} = (z_e +1)\,\frac{\Delta v_0}{c}\,,
\end{equation}
where $\Delta v_0$ is the passband in velocity units, and the last two equalities are asymptotically correct for $\Delta \nu_0\ll \nu_0$.

All our calculations were made with the fiducial value of $\Delta v_0 = 10^4\,$\kms\ ($\Delta\nu_0/\nu_0 = 0.033$), for five values of $f_{th}$: 0, 0.1, 1, 10, and 100 $\mu$Jy. Setting the value of $f_{th}$, we have the code loop over the grid values of $\nu_0$. For each grid value, the code loops over a range of principal quantum numbers ($5\le n \le 1000$) to find for each H$n\alpha$ line the values of $z_e(\nu_0,n)$ and $|\Delta z_e|\,(z_e)$, from equations~(1) and (4), respectively. Then the code loops over a grid of the absolute AB magnitudes of the galaxies at 1500\AA, $M_{AB}(1500)$, to find the number of galaxies, within $|\Delta z_e|$, that produce at least one detectable H$n\alpha$ line each. The total number of detectable galaxies, $N_{det}(\nu_0)$, determines the minimum search area $\Omega_4(\nu_0)= 4\pi/N_{det}(\nu_0)$. The index ``4'' indicates that calculations were done for $\Delta v_0 = 10^4\,$\kms.

The relation between $M_{AB}(1500)$ of a galaxy and the expected flux density at the center of the H$n\alpha$ line from this galaxy, $f_{{\rm H}n\alpha}$, is found as follows.

The line flux density is:
\begin{equation}
f_{{\rm H}n\alpha} = \dot N_{{\rm H}n\alpha}\,\frac{h\nu_{{\rm H}n\alpha}}{\delta\nu_{{\rm H}n\alpha}}\, \frac{(z+1)}{4\pi d_L^2} = \epsilon_{{\rm H}n\alpha}V\,\frac{hc}{\delta v}\, \frac{(z+1)}{4\pi d_L^2}\;,
\end{equation}
where $\dot N_{{\rm H}n\alpha}$ is the total rate of production of H$n\alpha$ photons in the galaxy; $\epsilon({\rm H}n\alpha)$ is the photon emissivity (${\rm cm}^{-3}\,{\rm s}^{-1}$), $V$ is the volume of the emitting gas; $\nu_{{\rm H}n\alpha}$ is the rest-frame frequency of the line; $\delta\nu_{{\rm H}n\alpha}$ and $\delta v$ are the widths of the line in frequency and velocity units, respectively (the latter being the same for all the lines), $d_L$ is the luminosity distance to the source, and the factor $(z + 1)$ accounts for the shrinkage of the observed frequency band as compared with the emitted frequency band. For a flat Universe ($1-\Omega_m - \Omega_{\Lambda} = 0$), the luminosity distance is
\begin{equation}
d_L = \frac{c}{H_0}\,(z+1)\,\int_0^z{\frac{{\rm d}z^{\prime}}{[\Omega_m\,(z^{\prime}+1)^3 + \Omega_{\Lambda}]^{1/2}}}\,,
\end{equation}
where $H_0$ is the Hubble constant, and $\Omega_m$ and $\Omega_{\Lambda}$ are the total matter density and the dark energy density (in the units of critical density), respectively. The luminosity distance was calculated by numerical integration, adopting the WMAP 9 $\Lambda$CMD cosmological parameters: $H_0 = 70\,$\kms, $\Omega_m = 0.28$ and $\Omega_{\Lambda} = 0.72$ \citep{2012arXiv1212.5226H}.

According to numerous calculations of hydrogen level populations in $H^+$ regions (e.g. \citet{1995MNRAS.272...41S}), the ratio of $\epsilon_{{\rm H}n\alpha}$ to the total recombination rate, $\alpha_r N_e^2$, is practically a constant for a broad range of electron densities. Using Hummer \& Storey's results for Menzel's Case B (for which the total coefficient of photorecombination to the levels $n\ge 2$, $\alpha_r \approx 4\times 10^{-13}\, {\rm cm^3\, s^{-1}}$), we derived the following empirical equation:
\begin{equation}
\epsilon_{{\rm H}n\alpha} \approx 3.25\,n^{-2.72}\alpha_r N_e^2\;.
\end{equation}
This equation is based on the results for $\alpha$-transitions within the interval $n = 3$ to 29, where it seems to be accurate within a few percent for any value of $N_e$. With our goal of only rough estimates of expected fluxes, we used this equation for all the higher values of $n$, up to $n = 1000$.

The condition of ionization equilibrium requires
\begin{equation}
\alpha_r N_e^2 V \approx \dot N_{912}\,(1-f_{\rm esc}^{912})\;,
\end{equation}
where $\dot N_{912}$ is the rate of production of ionizing photons in the galaxy and $f_{\rm esc}^{912}$ is the escape fraction of these photons. The connection between $\dot N_{912}$ and the luminosity of the galaxy at $\lambda 1500\,$\AA\ can be obtained in two steps. First, the $\lambda$1500\AA\ luminosity is related to the star formation rate (SFR) through \citep{2011ApJ...729...99M}:
\begin{equation}
\frac{\rm SFR}{\rm M_{\odot}\,yr^{-1}} \approx \frac{L_{1500}}{2\cdot 10^{28}\; {\rm erg\,s^{-1}\,Hz^{-1}}}\;.
\end{equation}
Second, with some assumptions about the initial mass function and the metallicity of the stars, one can obtain a relation between the SFR and the production rate of ionizing photons. Based on Table~4 from \citet{2003A&A...397..527S}, we can adopt for the first-generation low-metallicity stars:
\begin{equation}
\frac{\dot N_{912}}{10^{54}\,\rm photon\,s^{-1}} \approx \frac{\rm SFR}{\rm M_{\odot}\,yr^{-1}}\;.
\end{equation}
The last two equations are combined to yield
\begin{equation}
\frac{\dot N_{912}}{\rm photon\,s^{-1}}\approx 5\times 10^{25} \frac{L_{1500}}{ {\rm erg\,s^{-1}\,Hz^{-1}}}\,.
\end{equation}
From the standard definition of the absolute magnitude, M($\nu$), and the definition of the observed monochromatic AB magnitude, $m_{AB}(\nu)$,
\begin{equation}
{\rm log}_{10}f_{\nu} = \frac{-m_{AB}(\nu) - 48.6}{2.5}\;,
\end{equation}
where $f_{\nu}\,({\rm erg\,cm^2\,s^{-1}\,Hz^{-1}})$ is the observed flux density, one gets the relation between the absolute AB magnitude of the source and its monochromatic luminosity:
\begin{equation}
L_{\nu} = 4\pi\,(3.086\times 10^{19})^2\,f_{\nu}[M_{AB}(\nu)] = 4.346\times 10^{20}\times 10^{- 0.4\,M_{AB}(\nu)}\;{\rm erg\,s^{-1}\,Hz^{-1}}\,.
\end{equation}

From equations~(5), (7), (8), (11), and (13), we finally get:
\begin{equation}
f_{{\rm H}n\alpha} \approx 1.2\times 10^{-7}\, n^{-2.72}\, 10^{- 0.4\,M_{AB}^{1500}}(z+1)\left (\frac{d_L}{10^5\,{\rm Mpc}}\right )^{-2}\,\left(\frac{\delta v}{100\,{\rm km\, s^{-1}}}\right )^{-1} (1 - f_{\rm esc}^{912})\;({\rm\mu Jy}).
\end{equation}

The comoving number density of galaxies as a function of $z$ and $M_{AB}$ is found from the Schechter luminosity function (we drop the subindex $AB$):
\begin{equation}
\phi(M,z)\;({\rm Mpc^{-3}\,mag^{-1}}) = (0.4\,{\rm ln}10)\, \phi^*(z)\,\{10^{0.4[M^*(z) - M]}\}^{1+\alpha(z)}\,{\rm exp}\,\{-10^{0.4[M^*(z) - M]}\}\;.
\end{equation}
The values of the parameters $\phi^*(z)$, $M^*(z)$, and $\alpha(z)$ are calculated using the empirical equations based on the values of these parameters for 10 discrete values of $z$ (0.3, 1, 2,..., 8, 10) in the UV luminosity model of galaxies by \citet{2013ApJ...768L..37T}, which is in a good agreement with the observed statistics of galactic luminosities. The empirical equations were determined separately for $0\leq z <5$ and $5\leq z \leq 10$:
\begin{equation}
\phi^*_{-3} \approx -0.08251\,z^5 + 1.2092\,z^4 - 6.5396\,z^3 + 15.892\,z^2 - 16.880\,z + 8.0006\; (0\leq z <5)\,,
\end{equation}
\begin{equation}
\phi^*_{-3} \approx 14.316\,{\rm e}^{-0.418\,z}\; (5\leq z \leq 10)\,;
\end{equation}
\begin{equation}
M^* \approx -0.711\,{\rm ln}\,z - 19.842\; (0\leq z < 5)\,,
\end{equation}
\begin{equation}
M^* \approx 1.1683\,{\rm ln}\,z - 22.501\; (5\leq z \leq 10)\,;
\end{equation}
\begin{equation}
\alpha \approx -0.155\,{\rm ln}\,z - 1.5239\;(0\leq z < 5)\,,
\end{equation}
\begin{equation}
\alpha \approx -0.595\,{\rm ln}\,z - 0.7358\;(5\leq z \leq 10)\,.
\end{equation}
Note that the first two equations give $\phi^*_{-3}$ in the units of $10^{-3}\,$Mpc$^{-3}\,$mag$^{-1}$; the parameter $\phi^*$ entering equation~(15) is obtained from $\phi^*_{-3}$ by dividing by $10^3$. 

\section{RESULTS AND DISCUSSION}

\subsection{Blind Search}

From equation~(14), for the most luminous galaxies ($M_{AB}^{1500} \sim -22$) at $z\sim 10$ ($d_L \sim 10^5\,$Mpc), assuming $f_{esc}^{912} = 0$, the expected flux densities in the $n\sim 10$ lines are $\sim 1\; \mu$Jy, and for such galaxies at $z\simless 1$ ($d_L \sim 10^3\,$Mpc) the fluxes in the $n\sim 10$ lines are $\sim 10^2\;\mu$Jy. Modern radio astronomy is at the threshold of mastering the $\mu$Jy level. However, the galaxies of $M_{AB}^{1500} \sim -22$ must be very rare, especially at redshifts $\simgreat 10$. To assess the prospect of detecting {\it any} galaxies  in a blind search via their high-$n$ hydrogen recombination lines we used the approach described in \S~2. We assumed that the width of all the lines is $\Delta v = 100\;$\kms\ and that $f_{esc}^{912} = 0$.

The results of the calculations are presented in Figures~1 and 2.
\begin{figure}
\centering
\includegraphics[scale=0.9]{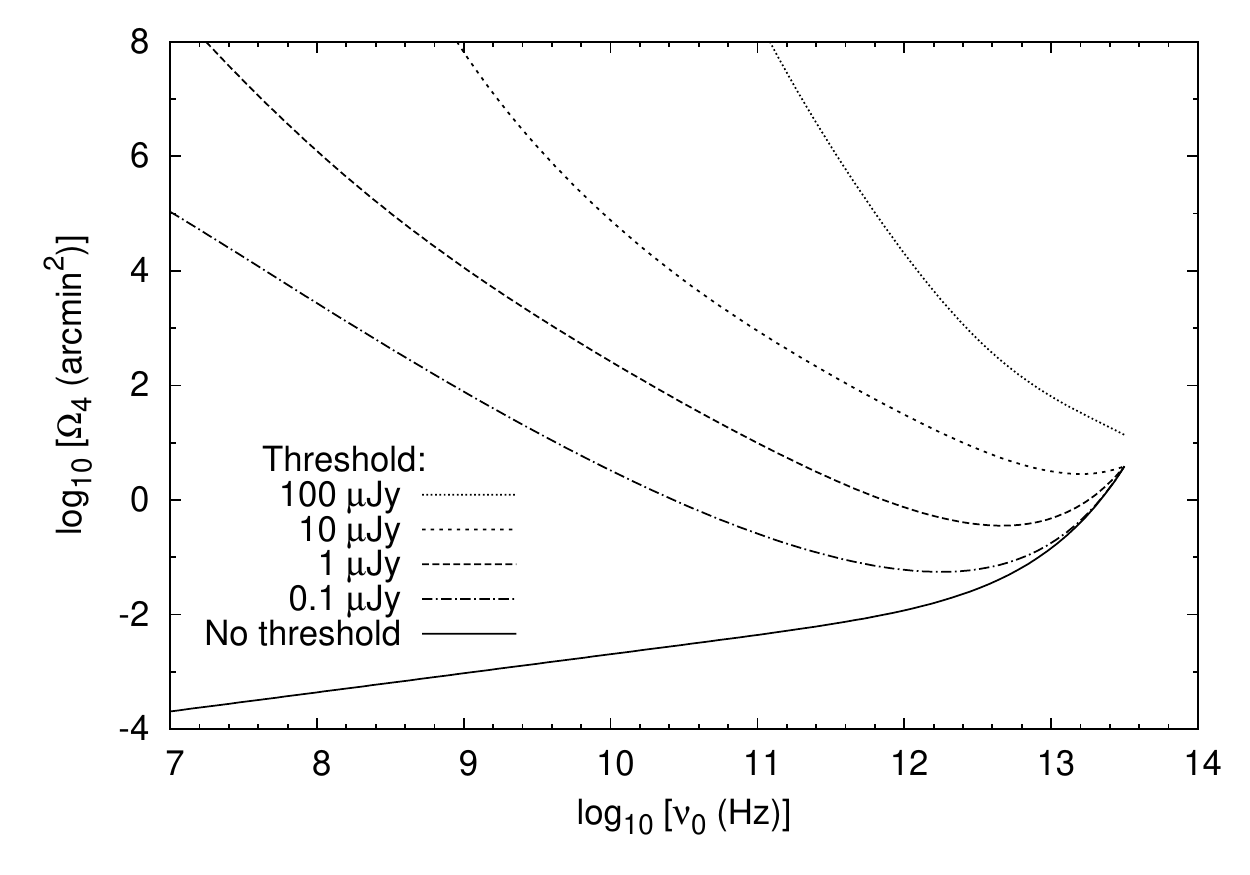}
\caption{The minimum sky area within which at least one detectable galaxy is expected from a redshift $z \le 10$, as a function of the observing frequency and the threshold of detection.}
\label{fig:fig1}
\end{figure}
\begin{figure}
\centering
\includegraphics[scale=0.9]{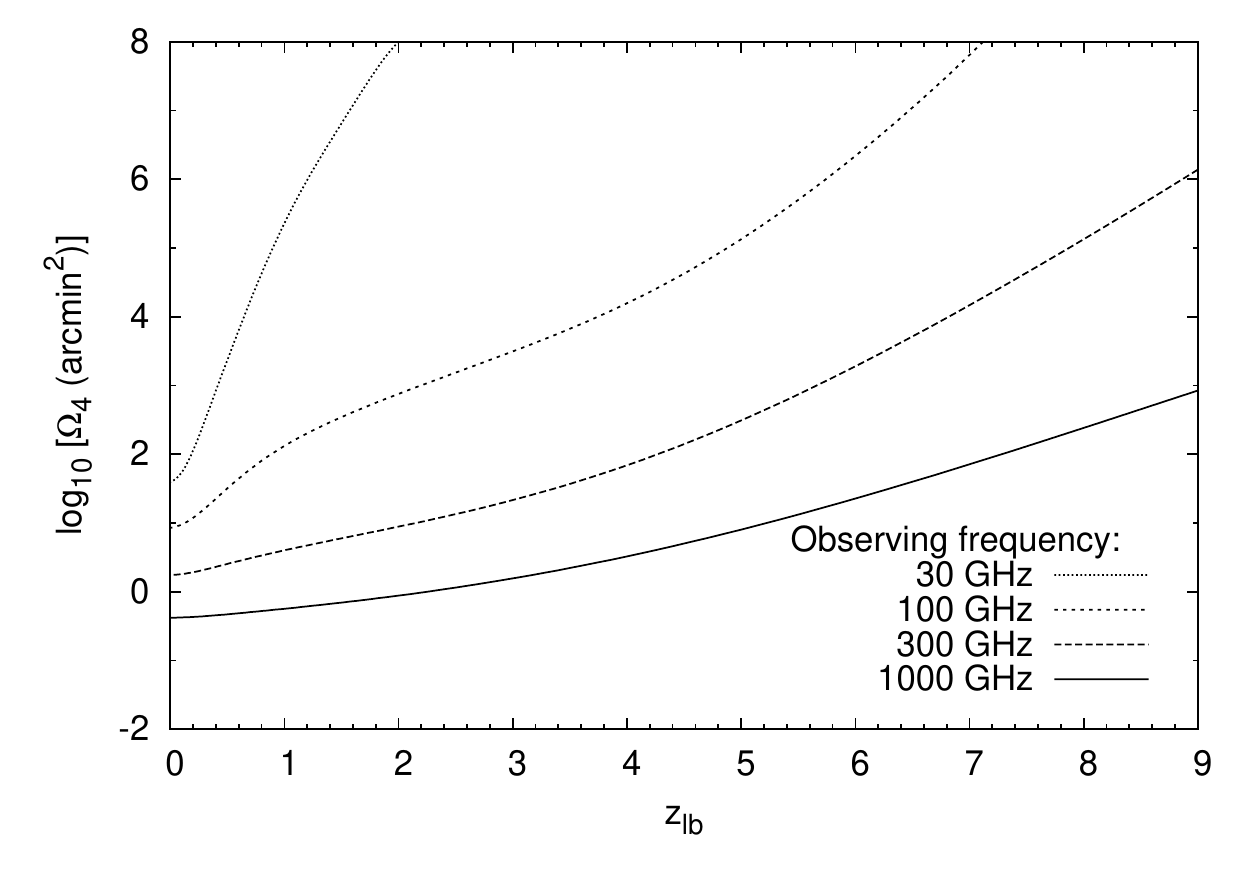}
\caption{The minimum sky area within which at least one detectable galaxy is expected from the redshift range between the lower boundary $z = z_{lb}$ and $z =10$, as a function of $z_{lb}$ and $\nu_0$, for the detection threshold of 1~$\mu$Jy.}
\label{fig:fig2}
\end{figure}
Figure~1 shows the dependence of $\Omega_4$ on the observing frequency. The ``No threshold'' curve shows the total number of the star-forming galaxies that have any lines in the range of $n$ we consider ($5 \le n \le 1000$), which would fall in the passband $\Delta\nu_0 = 0.033\nu_0$ around an observing frequency $\nu_0$. The increase of the minimum search area with increasing detection threshold is understandable. The shape of individual curves, corresponding to various values of the detection threshold, can be explained by the combined effects of the dependence of $f_{{\rm H}n\alpha}$ on $n$; the dependence of the luminosity function on the redshift; the decreasing interval between the adjacent lines with the increasing $n$; and the limited range of the lines included in the calculations. For all the considered non-zero detection thresholds, the minimum search area decreases with increasing frequency, up to the highest radio frequency considered here, $\nu_0 \approx 1\,$THz.

As illustrative examples, we consider the prospects of a blind search with ALMA, SKA and LOFAR.

When fully operational, ALMA will be capable of observing from 84 to 950 GHz. Let us consider, for instance, a search at 100 GHz (ALMA's band 3). According to the ALMA sensitivity calculator, observations with a spectral resolution of 100 \kms\ and the maximum recoverable scale of 8.6 arcsec (the ALMA C32-6 configuration) will provide a detection threshold of $\sim 2\,\mu$Jy in one hour of integration. From Figure~2, the minimum search area for this threshold at $\nu_0 = 100\,$GHz is $\approx 20\,$arcmin$^2$. This area contains $\approx 1000\,$ elements of 8.6 arcsec scale. Thus, the total search time should be $\simgreat 1000\,$hours.

The prospect of detection is similar for SKA. If a sensitivity of $\sim 10^4\,{\rm m^2\,K^{-1}}$ is achieved for the planned highest frequency of 30 GHz, then, with the system's noise temperature $\sim 10\,$K, a detection threshold $\sim 10\,\mu$Jy will be achieved in one hour of integration for a spectral resolution of $\sim 100\,$\kms. According to Figure~2, this threshold corresponds to a minimum search area $\sim 10^5\,$arcmin$^2$, which is about three orders of magnitude larger than the expected instantaneous SKA field of view at this frequency. Thus, again, integration times $\simgreat 1000\,$hours will be needed for a detection.

According to www.astron.nl/radio-observatory/astronomers/lofar-imaging-capabilities-sensitivity/sensitivity-lofar-array/sensiti, LOFAR's sensitivity, calculated for a bandwidth of 3.66 MHz and 8 hours of integration, is $\approx 4\,$mJy at the lowest frequency of 30 MHz and $\approx 1\,$mJy at the highest frequency of 240 MHz. Recalculating for a 100 \kms\ bandwidth and 1000 hours of integration (to compare with the cases of ALMA and SKA), gives the sensitivities of $\approx 7\,$mJy and $\approx 0.6\,$mJy, respectively. Addressing Figure~1, we find that, for both frequencies, the minimum search area is much greater than the whole celestial sphere, which makes the search unrealistic.

These pessimistic prospects of detecting an optically thin spontaneous emission in high-$n$ lines from far-away galaxies in a blind search are made worse by the fact that the number of detectable galaxies drops with redshift. Figure~2 illustrates this fact. It shows the minimum search area for the detection threshold of 1 $\mu$Jy as a function of the redshift interval, which is set by its lower boundary $z_{lb}$, the upper boundary being fixed at $z=10$. It is seen that the minimum search areas derived from Figure~1 are determined by low-redshift galaxies. Detecting galaxies at higher redshifts with a good probability requires larger search areas. This requirement gets stricter with decrease of observing frequency. In particular, it will make the blind detection of the first generation galaxies at $z\sim 10$ with SKA, even at its  highest planned working frequency of 30 GHz, highly challenging, as long as the radiation is spontaneous and optically thin. 

\subsection{Photometrically Identified High-\texorpdfstring{$z$}{z} Galaxies}

Some of the photometrically identified (Lyman-break) bright galaxies at $z>5$ exceed the 1500A flux densities expected for their estimated redshift and luminosity, because of gravitational lensing. Detection of hydrogen recombination lines from such galaxies would allow for a precise determination of their redshifts. As an example, consider the lensed, $z\approx 6-7$, galaxy identified by \citet{2012Natur.489..406Z}. Its $m_{AB} \approx 23.9$ corresponds to $M_{AB} \approx -25.1$ at $z=6.5$ (the luminosity distance $d_L \approx 63,000\,$Mpc). According to equation~(14), the spontaneously emitted 100 \kms\ wide H$20\alpha$ and H$21\alpha$ lines from such a galaxy (the observing frequencies $\approx 102$ and 88~GHz, respectively) should yield a flux density of $\approx 6-7\,\mu$Jy. This will secure a $S/N \approx 5$ in $\approx\,3\,$hours of integration with the ALMA C32-6 configuration. Reversing equation~(4), we obtain the observing frequency uncertainty $|\delta\nu_0|$ corresponding to the redshift uncertainty $|\delta z_e|$:
\begin{equation}
\frac{|\delta\nu_0|}{\nu_0} = 2\, \left [\sqrt{\left (\frac{z_e+1}{|\delta z_e|} \right )^2 +1} - \left (\frac{z_e+1}{|\delta z_e|}\right ) \right ] \approx \frac{|\delta z_e|}{z_e+1}\,.
\end{equation}
The approximate equality corresponds to $|\delta z_e| \ll (z_e+1)$. Substituting $\nu_0 \approx 90-100\,$GHz, $z_e \approx6.5$, and $|\delta z_e| \approx 1$, we get $|\delta\nu_0|\approx 13\,$GHz. This matches the maximum available instantaneous passband of 16 GHz, which makes it possible, in principle, to detect $both$ lines within the same passband in a few hours of integration. 

\subsection{Adjustments due to non-Zero \texorpdfstring{$f_{\rm esc}^{912}$}{fesc912}, Dust Absorption, and Maser Amplification}

The above estimates of the flux densities due to optically thin spontaneous emission in high-$n$ hydrogen recombination lines may be affected by three factors: (1) the non-zero value of the escape fraction of ionizing photons, $f_{\rm esc}^{912}$, in equation~(14); (2) absorption of the line radiation by dust within the galaxy; and (3) maser amplification.

The value of $f_{\rm esc}^{912}$ is unknown, but as long as it is much smaller than unity, as observed at low redshifts \citep{2000ApJ...545...86W}, the escape of the ionizing photons would not affect significantly our estimates.

The most prospective, $n\sim 10$, transitions have the rest wavelengths $\sim 10^2\;\mu$m. Assuming, by analogy with the Milky Way, that the typical high values of the dust optical depth for the visible light are $\sim 10$ in the galactic plane or in dense dusty cocoons surrounding H$^+$ regions, and adopting the inverse power law for the drop of the dust absorption efficiency with wavelength, the optical depth for $n\sim 10$ lines must be $\ll 1$. Thus, we do not expect the dust to affect significantly the $n\sim 10$ recombination line luminosity of high-$z$ galaxies either.

Based on the analysis of maser saturation by \citet{1996ApJ...470.1118S}, we estimate that possible maser amplification in any of the $n\sim 10$ lines is limited by a factor $\sim 10$. So, even if masing does occur in some of these lines, it may improve the prospects of detection only slightly.

\section{CONCLUSIONS}

We studied the prospects of detecting high-$n$ hydrogen recombination lines from the first-generation galaxies at $z\simless 10$ with existing or forthcoming radio-astronomical facilities. A blind search for such lines seems to be challenging even with the best existing and planned facilities, such as ALMA, LOFAR, and SKA.  However, some $z\gg 1$ galaxies with photometrically estimated redshifts may be detectable in $n\sim 10$ lines with facilities like ALMA, if they are amplified by gravitational lensing. The detection of radio recombination lines from such galaxies would allow for a precise determination of their redshifts.

\section{ACKNOWLEDGEMENTS}

E.R. was a Maria Mitchell Observatory REU intern while working on this project. He gratefully acknowledges the support by the NSF REU grant AST-0851892 and by the Nantucket Maria Mitchell Association. The work was also supported in part by NSF grant AST-0907890 and NASA grants NNX08AL43G and NNA09DBB30A (for A.L.). The authors thank the anonymous referee for the careful reading of the paper and several valuable suggestions.

\bibliography{references}

\begin{thebibliography}{}
\expandafter\ifx\csname natexlab\endcsname\relax\def\natexlab#1{#1}\fi

\bibitem[{{Anantharamaiah} {et~al.}(2000){Anantharamaiah}, {Viallefond},
  {Mohan}, {Goss}, \& {Zhao}}]{2000ApJ...537..613A}
{Anantharamaiah}, K.~R., {Viallefond}, F., {Mohan}, N.~R., {Goss}, W.~M., \&
  {Zhao}, J.~H. 2000, \apj, 537, 613

\bibitem[{{Bouwens} {et~al.}(2011){Bouwens}, {Illingworth}, {Oesch},
  {Labb{\'e}}, {Trenti}, {van Dokkum}, {Franx}, {Stiavelli}, {Carollo},
  {Magee}, \& {Gonzalez}}]{2011ApJ...737...90B}
{Bouwens}, R.~J., {Illingworth}, G.~D., {Oesch}, P.~A., {et~al.} 2011, \apj,
  737, 90

\bibitem[{{Finkelstein} {et~al.}(2012){Finkelstein}, {Papovich}, {Ryan},
  {Pawlik}, {Dickinson}, {Ferguson}, {Finlator}, {Koekemoer}, {Giavalisco},
  {Cooray}, {Dunlop}, {Faber}, {Grogin}, {Kocevski}, \&
  {Newman}}]{2012ApJ...758...93F}
{Finkelstein}, S.~L., {Papovich}, C., {Ryan}, R.~E., {et~al.} 2012, \apj, 758,
  93

\bibitem[{{Gordon} \& {Sorochenko}(2002)}]{2002ASSL..282.....G}
{Gordon}, M.~A., \& {Sorochenko}, R.~L., eds. 2002, Astrophysics and Space
  Science Library, Vol. 282, {Radio Recombination Lines. Their Physics and
  Astronomical Applications}

\bibitem[{{Hinshaw} {et~al.}(2012){Hinshaw}, {Larson}, {Komatsu}, {Spergel},
  {Bennett}, {Dunkley}, {Nolta}, {Halpern}, {Hill}, {Odegard}, {Page}, {Smith},
  {Weiland}, {Gold}, {Jarosik}, {Kogut}, {Limon}, {Meyer}, {Tucker}, {Wollack},
  \& {Wright}}]{2012arXiv1212.5226H}
{Hinshaw}, G., {Larson}, D., {Komatsu}, E., {et~al.} 2012, ArXiv e-prints,
  arXiv:1212.5226

\bibitem[{{Loeb} \& {Furlanetto}(2012)}]{loeb12}
{Loeb}, A., \& {Furlanetto}, S.~R., eds. 2012, {The First Galaxies in the
  Universe}, Princeton Series in Astrophysics (Princeton University Press)

\bibitem[{{Mu{\~n}oz} \& {Loeb}(2011)}]{2011ApJ...729...99M}
{Mu{\~n}oz}, J.~A., \& {Loeb}, A. 2011, \apj, 729, 99

\bibitem[{{Ouchi} {et~al.}(2010){Ouchi}, {Shimasaku}, {Furusawa}, {Saito},
  {Yoshida}, {Akiyama}, {Ono}, {Yamada}, {Ota}, {Kashikawa}, {Iye}, {Kodama},
  {Okamura}, {Simpson}, \& {Yoshida}}]{2010ApJ...723..869O}
{Ouchi}, M., {Shimasaku}, K., {Furusawa}, H., {et~al.} 2010, \apj, 723, 869

\bibitem[{{Pritchard} \& {Loeb}(2012)}]{2012RPPh...75h6901P}
{Pritchard}, J.~R., \& {Loeb}, A. 2012, Reports on Progress in Physics, 75,
  086901

\bibitem[{{Schaerer}(2003)}]{2003A&A...397..527S}
{Schaerer}, D. 2003, \aap, 397, 527

\bibitem[{{Storey} \& {Hummer}(1995)}]{1995MNRAS.272...41S}
{Storey}, P.~J., \& {Hummer}, D.~G. 1995, \mnras, 272, 41

\bibitem[{{Strelnitski} {et~al.}(1996){Strelnitski}, {Ponomarev}, \&
  {Smith}}]{1996ApJ...470.1118S}
{Strelnitski}, V.~S., {Ponomarev}, V.~O., \& {Smith}, H.~A. 1996, \apj, 470,
  1118

\bibitem[{{Tacchella} {et~al.}(2013){Tacchella}, {Trenti}, \&
  {Carollo}}]{2013ApJ...768L..37T}
{Tacchella}, S., {Trenti}, M., \& {Carollo}, C.~M. 2013, \apjl, 768, L37

\bibitem[{{Wood} \& {Loeb}(2000)}]{2000ApJ...545...86W}
{Wood}, K., \& {Loeb}, A. 2000, \apj, 545, 86

\bibitem[{{Zheng} {et~al.}(2012){Zheng}, {Postman}, {Zitrin}, {Moustakas},
  {Shu}, {Jouvel}, {H{\o}st}, {Molino}, {Bradley}, {Coe}, {Moustakas},
  {Carrasco}, {Ford}, {Ben{\'{\i}}tez}, {Lauer}, {Seitz}, {Bouwens},
  {Koekemoer}, {Medezinski}, {Bartelmann}, {Broadhurst}, {Donahue}, {Grillo},
  {Infante}, {Jha}, {Kelson}, {Lahav}, {Lemze}, {Melchior}, {Meneghetti},
  {Merten}, {Nonino}, {Ogaz}, {Rosati}, {Umetsu}, \& {van der
  Wel}}]{2012Natur.489..406Z}
{Zheng}, W., {Postman}, M., {Zitrin}, A., {et~al.} 2012, \nat, 489, 406

\end{thebibliography}

\end{document}